\documentclass[aps,prl,groupedaddress,fleqn]{revtex4}
\pdfoutput=1
\usepackage{graphicx}
\usepackage{amsmath}
\begin{document}

\title{Non-perturbative treatment of strongly-interacting fields: insights from liquid theory}
\author{K. Trachenko$^{1}$}
\author{V. V. Brazhkin$^{2}$}
\address{$^1$ School of Physics and Astronomy, Queen Mary University of London, Mile End Road, London, E1 4NS, UK, email k.trachenko@qmul.ac.uk}
\address{$^2$ Institute for High Pressure Physics, RAS, 142190, Moscow, Russia}

\begin{abstract}
We outline a new programme of solving the problem of treating strong interactions in field theories. The programme does not involve perturbation theories and associated problems of divergences. We apply our recent idea of treating strongly interacting liquids to field theories by showing the equivalence of Hamiltonians of liquids and interacting fields. In this approach, the motion of the field results in the disappearance of $n-1$ transverse modes with frequency smaller than the Frenkel frequency $\omega_{\rm F}$, similar to the loss of two transverse modes in a liquid with frequency $\omega<\omega_{\rm F}$. We illustrate the proposed programme with the calculation of the energy and propagator, and show that the results can not be obtained in perturbation theory to any finite order. Importantly, the Frenkel energy gap $E_{\rm F}=\hbar\omega_{\rm F}$ and the associated massive Frenkel particle naturally appear in our consideration, the result that is relevant for current efforts to demonstrate a mass gap in interacting field theories such as Yang-Mills theory. Notably, our mechanism involves a physically sensible starting point in terms of real masses (frequencies) in the harmonic non-interacting field, in contrast to the Higgs effect involving the imaginary mass as a starting point. We further note that the longitudinal mode in our approach remains gapless, implying that both short-range and long-range forces with massive and massless particles naturally emerge and unify in a single interacting field, a result not hitherto anticipated. Finally, we comment on the relationship between our results and hydrodynamic description of the quark-gluon plasma.
\end{abstract}

\maketitle

\section{Introduction}

The frontiers of modern physics are often said to lie in two directions where no adequate theoretical description exists: micro-world (particle physics, unified description of interactions and so on) and mega-world (cosmological problems). On the other hand, theoretical description of the familiar macro-world is considered to be largely complete. This is not quite the case: whereas solids and gases are well understood, the third state of the macro-world, the liquid state, presents serious theoretical problems. Surprising though it may seem, these problems are similar to those existing in the micro-world and mega-world from the point of view of physics and mathematics.

The main problem of theoretical description of liquids is the absence of a small parameter: contrary to solids and gases, kinetic and potential energy of atoms in the liquid are comparable in magnitude. As a result, it is impossible to develop a theory of real strongly interacting liquids by starting from the gas state and turning the interaction on. It is equally impossible to develop a first-principles theory of liquids by starting from the harmonic solid and introducing anharmonicity of interactions which increase the amplitude of vibrations leading to melting because the problem of the macroscopic number of bifurcations in the non-linear many-body system is not tractable. Since using the small parameter approach and perturbation theory is the main method of theoretical physics, the ``gas'' description of liquids has been most popular despite its inadequacy. Yet more than half a century ago J Frenkel has proposed \cite{frenkel} a way to avoid infinitely complex mathematical problems involved in constructing a theory of liquids. He asserted that the motion of atoms in a liquid consists of almost harmonic oscillations as in the solid and fairly rare jumps of atoms between different adjacent equilibrium positions. The average time between these jumps is liquid relaxation time, $\tau$. Calculating $\tau$ from first principles still remains an impossible task, but if we obtain $\tau$ from another experiment or computer simulation we are able to derive and explain most important liquid properties. In the last few years, we have used this approach to construct a thermodynamic theory of liquids in the wide range of parameters and explain excitation spectra and other physical properties of liquids \cite{prb1,scirep1,phystoday}. Below we discuss how this approach can be applied to strong interactions in field theories and elementary particles.

In field theories, interactions are introduced to free field Hamiltonians to describe the processes of production of new particles, calculate their cross-sections as well as discuss interactions between different types of fundamental forces \cite{bogol,peskin,ryder}. Problems related to divergences and infinities originating in perturbation theories of interactions have been raised at the early stages of development of field theories \cite{dirac}, stimulating new ideas, tour-de-force calculations as well as deeper philosophical issues \cite{peskin}. The problem of divergences and divergent perturbation series remains profound and fundamental \cite{suslov}.

Notably, the treatment of interactions in field theories has been based on the premise that the only way of treating interaction terms is by using the perturbation theory (see, for example, p. 200 in Ref. \cite{ryder}). The divergences arising in perturbation expansions are therefore viewed as a common necessity \cite{bogol,peskin}.

Alongside with the mainstream perturbative approaches, non-perturbative aspects of quantum field theory were explored and discussed, including the problems involving strong interactions, or strong coupling (see, e.g., Refs. \cite{n1,n2,n3,n4,poli1,poli2,gubser}). These and other ideas were aimed, among other things, at finding ways around divergences and non-renormalizabilities of various types, and were successful in some respects but not in others \cite{n1,n2,n3,n4,n5}.

Here, we propose a new way of treating interactions that does not involve perturbation expansions and associated divergences. The new effective non-perturbative description is inspired by our recent theory of strongly-interacting liquids based on Frenkel reduction and relaxation time $\tau$ \cite{prb1,scirep1}.

In the next two Chapters, we review our recent theory of strongly interacting liquids and discuss the implications of these results for the first-principles Hamiltonian approaches. We subsequently show the equivalence of Hamiltonians describing strongly interacting liquids and interacting fields, enabling us to apply the recent theory of liquids to interacting fields. This specific proposal continues a more general trend of cross-fertilization between condensed matter physics and field theory (see, e.g. Ref. \cite{new9}).

In this approach, the motion of the field consists of oscillations in one harmonic well and hopping between different wells. This motion results in the disappearance of ($n$-1) modes with frequency smaller than the Frenkel frequency $\omega_{\rm F}=\frac{1}{\tau}$, similarly to the loss of two transverse modes in a liquid with frequency $\omega<\omega_{\rm F}$. We illustrate the proposed programme with the calculation of the energy and propagator of the strongly-interacting field, and show that the results can not be obtained in perturbation theory to any finite order.

Importantly, the Frenkel energy gap for transverse modes, $E_{\rm F}=\hbar\omega_{\rm F}$ and associated massive Frenkel particle naturally appear in our consideration, the result that is relevant for current efforts to demonstrate a mass gap in interacting field theories such as Yang-Mills theory. Notably, our mechanism involves a physically sensible starting point in terms of real masses (frequencies) in the harmonic non-interacting field, in contrast to the Higgs effect involving the imaginary mass as a starting point. We further note that the longitudinal mode remains gapless, implying that both short-range and long-range forces with massive and massless particles naturally emerge and unify in a single interacting field, a result not hitherto anticipated.

\section{Non-perturbative thermodynamic theory of strongly-interacting liquids}

Heat capacity of matter is widely perceived to be one of its main important properties because it informs a scientist about the degrees of freedom available in the system. Heat capacity is well understood in gases and solids but not in the third state of matter, the liquid state, despite over a century of intensive research.

In an amusing story about his teaching experience, Granato recalls living in fear about a potential student question about liquid heat capacity \cite{granato}. Observing that the question was never asked by a total of 10000 students, Granato proposes that ``...an important deficiency in our standard teaching method is a failure to mention sufficiently the unsolved problems in physics. Indeed, there is nothing said about liquids [heat capacity] in the standard introductory textbooks, and little or nothing in advanced texts as well. In fact, there is little general awareness even of what the basic experimental facts to be explained are.'' Granato's observation is supported by the absence of a discussion of liquid specific heat in classic statistical physics textbooks as well as texts dedicated to liquids and other disordered systems \cite{lanstat,ziman,l1,l2,l3,l4,l5}).

Perturbation approaches do not apply to real liquids because neither interactions nor virial parameter, or density, are small. For example, the specific heat of the Van Der Waals system with added interactions is $\frac{3}{2}k_{\rm B}$ \cite{lanstat}, as in the ideal gas. Due to the absence of a small parameter, perturbation approaches have not been considered as serious attempts to explain thermodynamic properties of real strongly-interacting liquids. The absence of a small parameter was, in Landau view, the fundamental property of liquids that ultimately precluded the construction of a theory of liquids at the same level existing for solids or gases. In a similarly discouraging spirit, Landau and Lifshitz Statistical Physics textbook states twice that strong interactions, combined with system-specific form of interactions, imply that liquid energy is strongly system-dependent, precluding the calculation of energy and other properties in general form, contrary to solids or gases \cite{lanstat}.

We note that strong interactions are successfully treated in solids in the phonon approach where the small parameter is atomic displacements and where a harmonic approximation to the energy is often a good approximation. This approach has long been thought to be inapplicable to liquids where atomic displacements are large. As a result, liquids have been thought to have no small parameter that immensely simplifies matters in solids and gases: in liquids displacements are large and interactions are strong at the same time.

We have recently proposed a way out of this seemingly hopeless situation by putting forward a non-perturbative phonon theory of liquids \cite{prb1,scirep1}. Here and below, by a ``non-perturbative'' approach we generally mean an approach that does not involve perturbation theory and expansions in terms of interaction, with reference to both liquids and interacting fields. Our approach is based on the concept of liquid relaxation time introduced by J. Frenkel \cite{frenkel}. Frenkel realized that in order to make progress, in particular in rationalizing the emerging experimental data at his time, one needs to move away from first-principles Hamiltonians that offered no way of solution for the reasons above, including those articulated by Landau. With a remarkable physical insight, Frenkel introduced liquid relaxation time $\tau$ as the average time between particle jumps at one point in space in a liquid, and subsequently showed that $\tau$ is related to liquid viscosity $\eta$ via the Maxwell relationship $\eta=G_{\infty}\tau$, where $G_{\infty}$ is the instantaneous shear modulus.

In the Frenkel picture, molecular dynamics in liquids consists of two types: solid-like oscillations around quasi-equilibrium positions and ballistic jumps between these positions taking place with a time period of $\tau$, the picture that has been subsequently confirmed in numerous experimental and modeling studies.

$\tau$ depends on liquid structure and interactions in a very complicated way, with arguably no feasible way of calculating it in a general case. This is because $\tau$ is governed by the activation barriers in the energy landscape which becomes intractably complex for a many-body system with arbitrary types of interactions and structural correlations. However, Frenkel's idea was that such a calculation is not required in order to understand the most basic features of liquid behavior. Lets start with the ability of liquids to maintain solid-like shear modes.

At times shorter than $\tau$, a liquid is a solid, and therefore supports one longitudinal mode and two transverse modes. At times longer than $\tau$, liquid flows and loses its ability to support solid-like transverse (shear) modes, and supports one longitudinal mode only as any elastic medium. This is equivalent to asserting that the only difference between a solid glass and a liquid is that a liquid does not support all solid-like transverse modes as the solid glass does, but only those with high frequency $\omega>\frac{1}{\tau}$. The longitudinal mode is unaffected in this picture: regardless of the flow, a liquid supports fluctuations of density as any elastic medium. In a dense liquid, these fluctuations extend to wavelengths comparable with the shortest interatomic separation as discussed below. The only modification that the flow process introduces to the longitudinal mode is that dissipation laws (namely, propagation ranges of harmonic waves) are different in the regimes $\omega>\frac{1}{\tau}$ and $\omega<\frac{1}{\tau}$ \cite{frenkel}.

Derived on purely theoretical grounds, Frenkel's idea was later experimentally confirmed, although with a significant time lag for high-frequency phonons that give the largest contribution to the energy. The ability of liquids to support solid-like collective modes with wavelengths extending to the shortest distance comparable to interatomic separations has been widely ascertained experimentally on the basis measured dispersion curves \cite{burkel,pilgrim}. This includes shear modes in both highly-viscous \cite{w3} and low-viscous liquids such as Na, Ga, water and so on \cite{mon-ga,mon-na,hoso,water}. Notably, most of this experimental evidence is fairly recent, and has started to come to the fore only when powerful synchrotron radiation sources started to be deployed, some 50--60 years after Frenkel's prediction. This long-lived absence of experimental data about propagating collective excitations in liquids may have contributed to their poor understanding from the theoretical point of view (see the recent review of this subject in Ref. \cite{review}).

In our solid-like approach to liquid thermodynamics \cite{prb1,scirep1}, we have taken up Frenkel's idea about liquid vibrational states and its ability to support high-frequency shear modes. The absence of two shear modes with frequency $\omega<\frac{1}{\tau}$ is equivalent to the potential energy of restoring forces for these modes becoming zero. Adding the energy of one longitudinal mode and the energy of diffusing atoms gives the liquid energy as

\begin{equation}
E=K+P_l+P_t(\omega>\omega_{\rm F})
\label{len}
\end{equation}
\noindent where $K$ is total kinetic energy that includes both oscillatory and ballistic components from jumping atoms ($\frac{3}{2}k_{\rm B}T$ per particle in the classical case), $P_l$ is the potential energy of the longitudinal mode, $P_t$ is the potential energy of two transverse modes with frequency $\omega>\omega_{\rm F}$, and we introduced the Frenkel frequency $\omega_{\rm F}=\frac{1}{\tau}$.

$E$ can be re-written in the form convenient for further calculations using the virial theorem, giving \cite{prb1,scirep1}

\begin{equation}
E=E_l+E_t(\omega>\omega_{\rm F})+\frac{1}{2}E_t(\omega<\omega_{\rm F})
\label{len1}
\end{equation}
\noindent where $E_l$ is the energy of the longitudinal mode and $E_t(\omega>\omega_{\rm F})$ is the energy of two transverse modes with frequency $\omega>\omega_{\rm F}$, respectively. The last term in Eq. (\ref{len1}) ensures that the loss of two transverse modes with frequency $\omega<\omega_{\rm F}$ is accompanied by the reduction of their potential energy only (as Eq. (\ref{len}) requires), but not by the reduction of the total kinetic energy of the system $K$.

$E_l$, $E_t(\omega>\omega_{\rm F})$ and $E_t(\omega<\omega_{\rm F})$ in Eq. (\ref{len1}) can be evaluated as $\int\limits_0^{\omega_{\rm D}}E(\omega,T)g_l(\omega)d\omega$, $\int\limits_{\omega_{\rm F}}^{\omega_{\rm D}}E(\omega,T)g_t(\omega)d\omega$ and $\int\limits_0^{\omega_{\rm F}}E(\omega,T)g_t(\omega)d\omega$, respectively, where $\omega_{\rm D}$ is Debye frequency (the largest oscillation frequency present in the system), $E(\omega,T)=\frac{\hbar\omega}{2}+\frac{\hbar\omega}{\exp\frac{\hbar\omega}{T}-1}$ (here and below, $k_{\rm B}=1$) and $g_l(\omega)$ and $g_t(\omega)$ are densities of states of longitudinal and transverse modes per atom, respectively: $g_l=\frac{3\omega^2}{\omega_{\rm D}^3}$ and $g_l=\frac{6\omega^2}{\omega_{\rm D}^3}$ \cite{prb1} (see Ref. \cite{review} for the recent review of the application of Debye model to liquids). This gives the energy per particle in the classical case as \cite{prb1}:

\begin{equation}
\epsilon=T\left(3-\left(\frac{\omega_{\rm F}}{\omega_{\rm D}}\right)^3\right)
\label{class}
\end{equation}
\noindent and in the quantum case as \cite{scirep1}:

\begin{equation}
\epsilon=\epsilon_0+T\left(3D\left(\frac{\hbar\omega_{\rm D}}{T}\right)-\left(\frac{\omega_{\rm F}}{\omega_{\rm D}}\right)^3D\left(\frac{\hbar\omega_{\rm F}}{T}\right)\right)
\label{enf}
\end{equation}
\noindent where $\epsilon_0=0$ is the zero-point energy, $D(x)=\frac{3}{x^3}\int\limits_0^x\frac{z^3{\rm d}z}{\exp(z)-1}$ is Debye function and where we omitted the anharmonic effects related to frequency softening.

The first term in brackets in Eqs. (\ref{class},\ref{enf}) is the familiar vibrational energy of the classical and quantum solid, respectively. The second term in Eqs. (\ref{class},\ref{enf}) describes the progressive loss of two transverse modes with frequency $\omega>\omega_{\rm F}=\frac{1}{\tau}$ on temperature increase in the liquid.

\begin{figure}
\begin{center}
{\scalebox{0.35}{\includegraphics{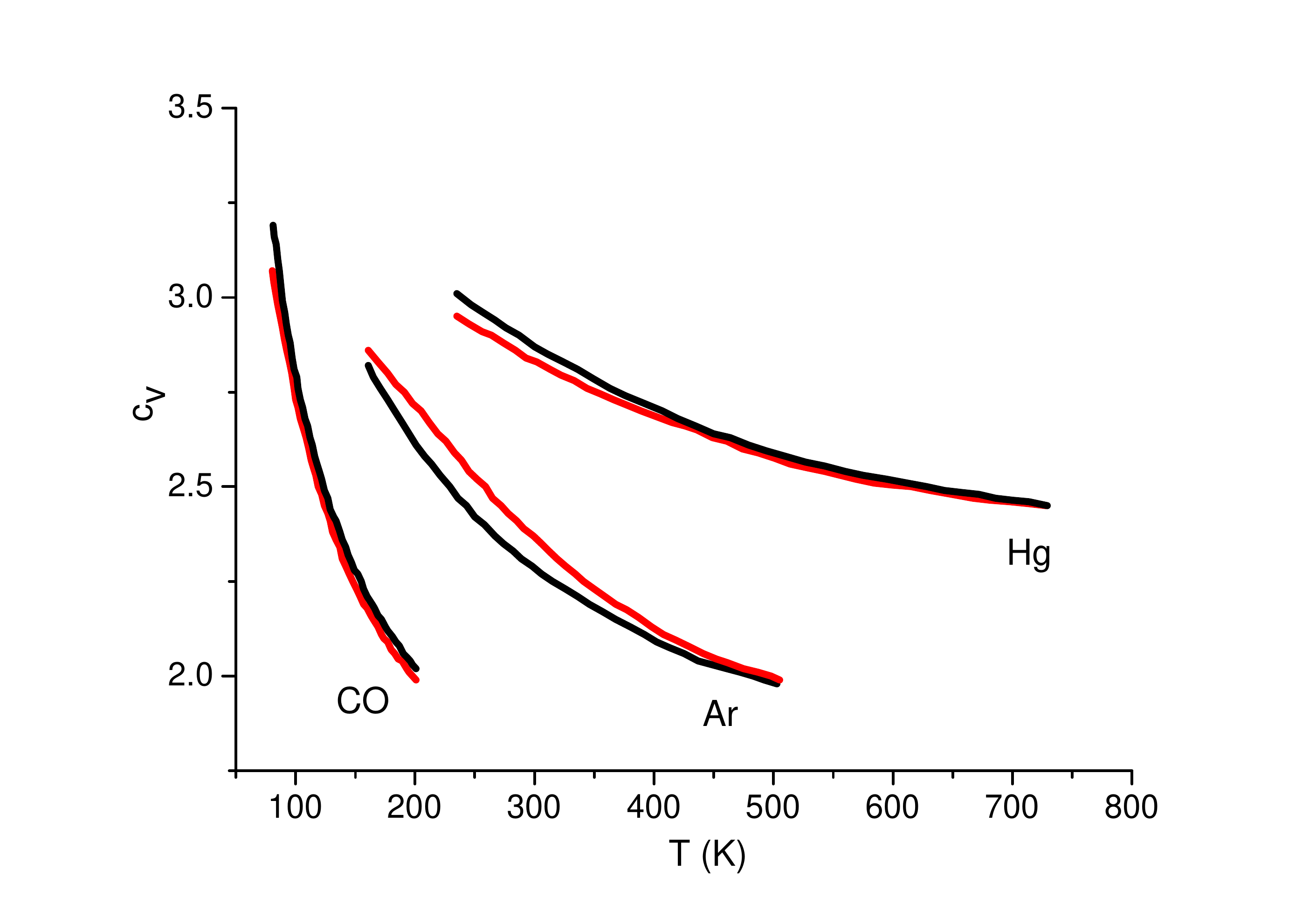}}}
\end{center}
\caption{Experimental (black color) and theoretical (red colour) $c_v$ ($k_{\rm B}=1$) for liquid Hg, Ar and CO \cite{scirep1}. Theoretical $c_v$ was calculated using Eq. (\ref{enf}) with the account of anharmonicity and thermal expansion.}
\label{cv3}
\end{figure}

Eqs. (\ref{class},\ref{enf}) provide a relationship between the liquid energy and $\omega_{\rm F}$ which can be calculated as $\omega_{\rm F}=\frac{1}{\tau}=\frac{G_\infty}{\eta}$, where $G_\infty$ is the infinite shear modulus and viscosity $\eta$ can be taken from independent experimental data. We have found that the calculated specific heat $c_v=\frac{d\epsilon}{d T}$ agrees reassuringly well with experimental data for 21 noble, metallic, molecular and network liquids in a wide range of temperature and pressure using no free fitting parameters \cite{scirep1}. In Figure \ref{cv3} we show experimental and calculated $c_v$ for systems representing three different classes of metallic, molecular and noble liquids. We observe that liquid $c_v$ decreases from approximately 3 at low temperature to 2 at high, the effect related to the progressive loss of two transverse modes with frequency $\omega>\omega_{\rm F}=\frac{1}{\tau}$ on temperature increase \cite{prb1,scirep1}.

At high temperature when $\tau\approx\tau_{\rm D}$, all transverse modes in the liquid are lost, giving $c_v=2$ according to Eq. (\ref{class}). At this point, atomic dynamics undergoes an important crossover from the oscillatory-ballistic as in a liquid to the purely ballistic as in a gas. We have recently proposed that this crossover defines a new line on a phase diagram \cite{phystoday,pre,prl}, the ``Frenkel'' line, crossing which results in qualitative changes of most important physical properties of the system, including diffusion, viscosity, speed of sound, thermal conductivity and, more recently, specific heat \cite{natcom}.

We conclude this section by making an important general assertion about liquid $c_v$ in the regime $\tau\gg\tau_{\rm D}$, where $\tau_{\rm D}=\frac{1}{\omega_{\rm D}}$ is the shortest, Debye, vibration period of about 0.1 ps. As recently discussed \cite{scirep2}, it can be rigorously shown that in this regime, liquid energy and specific heat are, to a very good approximation, equal to those in the solid, and that this result is consistent with liquid entropy exceeding solid entropy. This result is also obtained from Eqs. (\ref{class}) and (\ref{enf}): liquid energy and specific heat are equal to their solid-state values unless $\omega_{\rm F}\approx \omega_{\rm D}$. This picture therefore predicts that liquid $c_v$ in this regime is equal to the Dulong-Petit value (corrected for the effects of anharmonicity): $c_v=3$, consistent with experimental results (see Ref. \cite{scirep2} and Figure 1).

For practical purposes, $\tau\gg\tau_{\rm D}$ (or $\omega_{\rm F}\ll\omega_{\rm D}$) holds starting from $\tau\gtrsim 10\tau_{\rm D}$. Perhaps not widely recognized, the condition $\tau\approx 10\tau_{\rm D}$ holds even for low-viscous liquids such a liquid monatomic metals (Hg, Na, Rb and so on) and noble liquids such as Ar near their melting points \cite{scirep1,nist}, let alone for more viscous liquids such as room-temperature olive or motor oil, honey and so on. Importantly, the condition $\tau\gtrsim 10\tau_{\rm D}$ corresponds to almost the entire range of $\tau$ at which liquids exist, the fact that was not appreciated in earlier work on liquids. Indeed, on lowering the temperature, $\tau$ increases from its smallest limiting value of $\tau=\tau_{\rm D}\approx 0.1$ ps to $\tau\approx 10^{3}$ s where, by definition, a liquid forms a glass at the glass transition temperature. Here, $\tau$ changes by 16 orders of magnitude. Consequently, the condition $\tau\gg\tau_{\rm D}$ applies in the range $10^3-10^{-12}$ s, spanning 15 orders of magnitude of $\tau$. This constitutes almost entire range of $\tau$ where liquids exist as such.

\section{Implications for the Hamiltonian approach: Frenkel reduction}

We now discuss and interpret the result of the previous section in terms of the first-principles Hamiltonian of the problem.

As envisaged by Frenkel, the energy of the liquid consists of solid-like oscillatory motion around quasi-equilibrium positions interrupted by ballistic jumps after time $\tau$. This motion is described by the following Hamiltonian:

\begin{equation}
H_l=\sum_i\frac{m_i\dot{x_i}^2}{2}+\frac{1}{2}\sum_{ij}\frac{k_{ij}(x_i-x_j)^2}{2}+\sum_i V(x_i)
\label{hamilt}
\end{equation}
\noindent where $i$ runs over the sites of topologically disordered liquid structure.

The first two terms represent the familiar harmonic Hamiltonian giving the oscillatory motion. The last term is responsible for atomic jumps in the liquid once $V(x)$ has a form with two or more minima (see Fig. 2). $V(x)$ can be generally represented by any smooth function of $x$ including the polynomial function, with a general requirement that such a function has a set of minima closely located in energy and separated by distances $a$ comparable to interatomic separations and a characteristic value of activation barrier $U$. $U$ is completely defined by the function $V(x)$ and its parameters. Then, the activated jumps over barrier $U$ in the classical liquid set the value of Frenkel relaxation time $\tau$ as

\begin{figure}
\begin{center}
{\scalebox{0.35}{\includegraphics{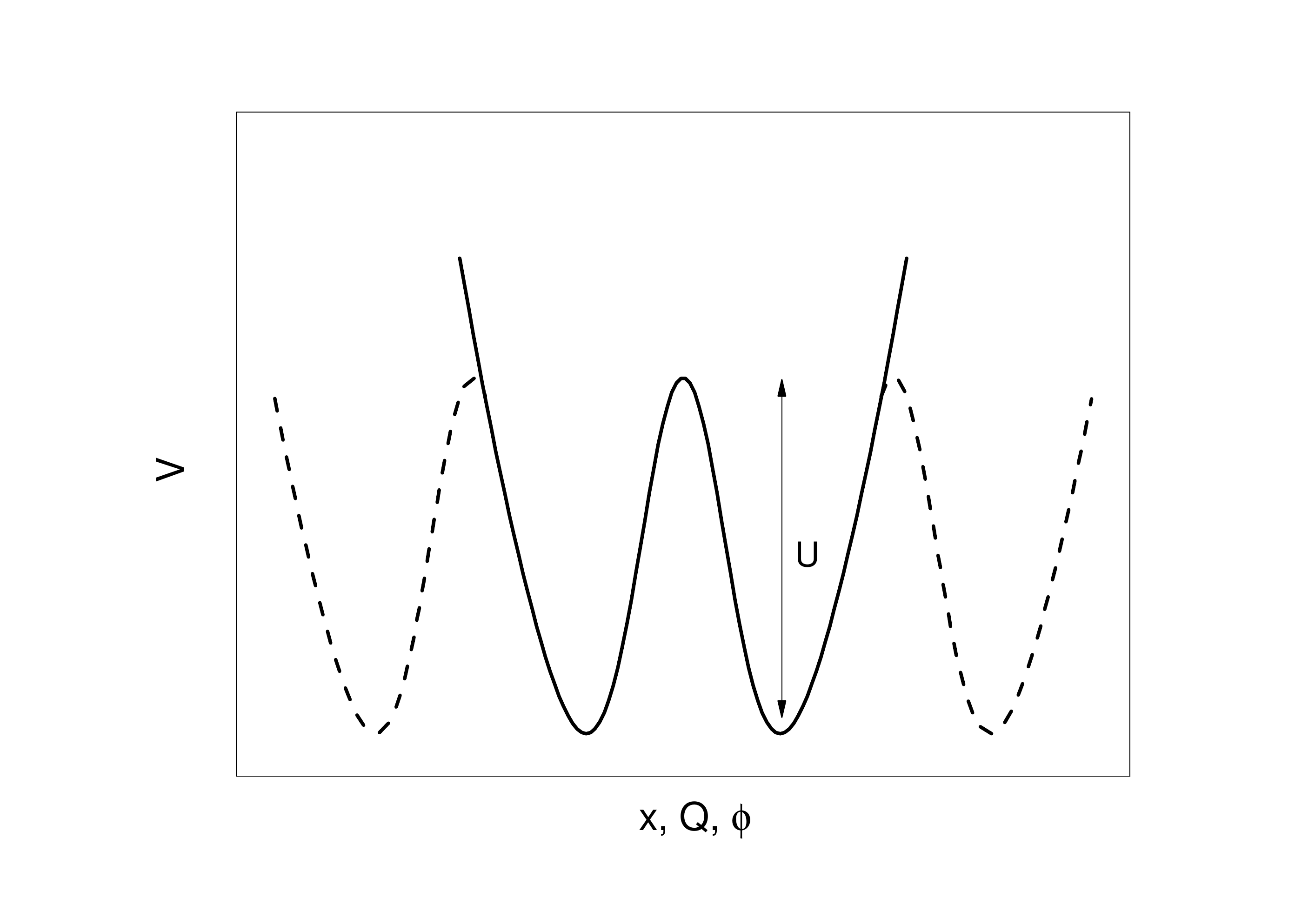}}}
\end{center}
\caption{Schematic representation of double-well (or multi-well) potential.}
\label{wells}
\end{figure}

\begin{equation}
\tau=\tau_{\rm D}\exp\left(\frac{U}{T}\right)
\end{equation}

We therefore find that Hamiltonian (\ref{hamilt}) describes liquid dynamics envisaged by Frenkel that includes oscillatory motion in one potential well and atomic jumps between adjacent wells with a period $\tau$.

The simplest analytical forms of $V(x)$ commonly considered include two higher powers of $x$:

\begin{equation}
V(x)=c_3x^3+c_4x^4
\label{3-4}
\end{equation}

Expansions similar to (\ref{3-4}) have been commonly used to model weak anharmonicity of interatomic interactions (see, e.g., Refs. \cite{marad,cowley}). Here, the important assumption was the smallness of $V(x)$, and corrections to the energy have been derived using the perturbation theory. A notable implication of the smallness of anharmonicity in this treatment is that $V(x)$ does not result in large atomic jumps between two possible minima of potential (\ref{3-4}) as is the case in liquids. Even with these simplifying assumptions, perturbative treatments of interaction (\ref{3-4}) involved extensive calculations which have nevertheless left several important open questions in the area \cite{marad,cowley,wallace,fultz,jpcm}.

$V(x)$ in Hamiltonian (\ref{hamilt}) is introduced to have the diagonal form because it straightforwardly endows the atoms with the possibility to jump between neighbouring positions during time $\tau$ as envisaged by Frenkel (see Figure 2). This corresponds to each atom moving in an external potential $V(x)$, whereas in reality $V(x)$ is set by the interactions between atoms that depend on relative distance between the atoms, $x_i-x_j$, as the second term in Eq. (\ref{hamilt}) does. Hence $V$ should contain non-diagonal higher powers such as $c_{ijk}x_ix_jx_k$, $c_{ijkl}x_ix_jx_kx_l$ \cite{marad,cowley} and so on. It will become apparent below that our main conclusion of the identity between the liquid Hamiltonian and interacting-field Hamiltonian does not depend on whether $V(x)$ retains its current diagonal form in Eq. (\ref{hamilt}) or depends on the cross product of different $x_i$.

It has proved impossible to calculate thermodynamic properties of general Hamiltonians such as (\ref{hamilt}) directly, by evaluating the partition function. Perturbative approaches such as those used to treat weak anharmonicity are not justified for real strongly-interacting liquids and for this reason have not been contemplated in the literature. Interestingly, this is in stark contrast to field theories where the perturbation approach has been the sole method of handling interactions, with infinite divergent series emergent \cite{bogol,peskin,ryder}. Even if the direct calculation involving the explicit treatment of $V(x)$ of liquid Hamiltonian (\ref{hamilt}) were possible, the results will be strongly dependent on $V(x)$ (i.e. system-dependent as discussed by Landau above), making the task of deriving general results for liquids seemingly hopeless. Yet this is not the main problem with attempting to directly evaluate the partition function with Hamiltonian (\ref{hamilt}).

In our view, the main problem with direct evaluation of the partition function based on Hamiltonians such as (\ref{hamilt}) is that the problem is largely under-specified, with no currently feasible general way of treating the ensuing complexity. Indeed, depending on the interaction term $V(x)$ and its parameters, Hamiltonian (\ref{hamilt}) can describe a dense strongly-interacting gas with ballistic motions only \cite{phystoday}, a liquid with both oscillations and ballistic jumps in between, or a solid with only weak anharmonicity of interactions where atomic jumps are so rare that they only occur at astronomical times and can be neglected \cite{prb2}. It is challenging to assert {\it a priori} which state of matter we are dealing with (\ref{hamilt}) because the answer depends on characteristic heights of activation barriers in a complex energy landscape of the problem, and calculating such a landscape for a general interacting many-body system is currently impossible. This is often illustrated as a story of some best physicists who are gathered on an island, asked to find the liquid state on the basis of given Hamiltonian and failing to find the liquid state despite being surrounded by water \cite{ash}.

Liquids therefore present an interesting and fundamental challenge of calculating the partition function on the basis of first-principles Hamiltonians, the challenge which is ultimately related to a complexity of the energy landscape which is not generally known. This problem is related to the deeper issue in statistical physics, that of performing calculations in restricted versus full phase space, including finding ways of accounting for important metastable states \cite{frenkel}. The problem is partially avoided by using simple model Hamiltonians that restrict the phase space as, for example, the harmonic Hamiltonian of a solid does.

The main idea behind our recent theory of liquids \cite{prb1,scirep1} is to move away from attempting to solve first-principle Hamiltonians. Instead, we are operating in terms of $\tau$ as the main physical property of the liquid that enables one to separate the atomic motion into oscillatory and jump motions \cite{scirep2,phystoday}. This has provided an enormous reduction of the problem (we call it ``Frenkel reduction'') because $\tau$ encompasses the physically relevant and important features of the unknown complex energy landscape. In this approach, we have been able to discuss liquid vibrational states, derive the final equation for the liquid energy, Eq. (\ref{enf}), and ascertain its agreement with experimental data.

Importantly, Frenkel reduction in terms of $\tau$ is applicable to any interacting non-linear system of many agents, which is intractable in general. The system of two coupled anharmonic oscillators with simple model forms of anharmonicity can be solved exactly, with the result that hopping between different equilibrium positions emerges as a result of a set of bifurcations \cite{nonlinear}. The barrier height $U$ (and hence $\tau$ in the case of thermal motion) can be related to the non-linearity parameters. This constitutes an important first-principles proof that hopping emerges as a result of bifurcations in the non-linear system. Unfortunately, increasing the number of agents or altering the simple analytical form of non-linearity makes the analysis impossible even using approximate schemes \cite{nonlinear}. Yet Frenkel reduction offers an important way in to analyze most important properties of the system as discussed in the previous section.

\section{Hamiltonians of liquids and interacting fields}

We now demonstrate the equivalence of liquid and interacting field Hamiltonians. This will enable us to apply the results derived for liquids in the previous section to interacting fields.

Lets consider field $\phi$ whose energy can be represented by a set of independent harmonic oscillators (normal modes) with frequencies $\omega_i$ as is the case for, for example, Klein-Gordon field. Below we consider $n$-component field and $n=3$ in particular to continue the analogy with liquids. We now add the ``interaction'' term $V(\phi)$ to the field Hamiltonian density, $H_f$, to arrive at the most common model of an interacting field \cite{bogol,peskin}:

\begin{equation}
H_f=\frac{1}{2}\sum_i\left(\dot{\phi_i}^2+\omega_i^2\phi_i^2+V(\phi_i)\right)
\label{field}
\end{equation}

We note that in condensed matter theory, ``interaction'' refers to interatomic forces, and includes both quadratic and higher anharmonic terms in Eq. (\ref{hamilt}). Strong interactions in liquids imply that both of these terms are not small compared to temperature: large quadratic terms imply that interatomic forces in liquids are as strong as in solids, and large anharmonic terms imply that these terms are not small compared to the quadratic terms, resulting in atomic hopping and liquid flow (as opposed to serving to merely modify the spectrum of a solid system due to anharmonicity). In field theories, ``interaction'' refers to terms higher than quadratic such as $V(\phi)$ in Eq. (\ref{field}). As in liquids, strong interactions here imply that these higher terms are not small, and are comparable to the quadratic terms.

Different types of $V(\phi)$ have been considered. Of these, most interesting are with two or more minima such as those shown in Figure 2, because these correspond to spontaneous symmetry breaking and the associated creation of massive fields and particles \cite{bogol}. Two-minima potentials have also been discussed in models of cosmology and evolution of the Universe \cite{kirzh}. Most simplest examples of $V(\phi)$ include:

\begin{equation}
\begin{aligned}
&V(\phi)=g\phi^4\\
&V(\phi)=-h\phi^3+g\phi^4\\
&V(\phi)=-g\phi^4+\lambda \phi^6\\
&...\\
\end{aligned}
\label{int}
\end{equation}

The first example above is the commonly studied $\phi^4$ model \cite{bogol,peskin,ryder} where the total potential has two minima if the quadratic potential term in Eq. (\ref{field}) is negative, $\omega_i^2<0$ \cite{bogol}, whereas in the second example the two minima can exist when $\omega_i^2>0$ and $h$ and $g$ are both positive. The third example of the double-well interaction was considered in our recent work \cite{scirep3} discussed below.

We note the difference between the liquid Hamiltonian, Eq. (\ref{hamilt}), and the interacting-field Hamiltonian, Eq. (\ref{field}): the first operates in terms of atomic coordinates $x$ whereas the second in terms of field variables $\phi$ which represent independent oscillators quantising the field, or normal modes. To bridge the difference, the first two terms in Eq. (\ref{hamilt}) could be written in terms of the normal modes $Q$, but the normal-mode transformation $x\rightarrow Q$ gives cross-terms in higher-order terms $V\propto Q_iQ_jQ_k...$ (by introducing these cross terms if $V(x)$ is initially chosen to be diagonal as in Eq. (\ref{hamilt}) or, if $V(x)$ is chosen to have cross terms $x_ix_jx_k...$ from the outset, by transforming the cross-terms of $x_i$ to cross-terms of $Q_i$). This might appear as a problem for establishing the equivalence between the liquid Hamiltonian and $H_f$ in Eq. (\ref{field}). Below we discuss an alternative method where this difficulty does not arise.

Lets consider a liquid Hamiltonian in the form similar to Eq. (\ref{field}), written in terms of normal modes $Q$ which includes an interaction term $V(Q)$ with at least two real minima \cite{scirep3}:

\begin{equation}
\begin{aligned}
&H(Q)=\frac{1}{2}\sum_i\left(\dot{Q_i}^2+\omega_i^2Q_i^2+H_{\rm int}(Q_i)\right)\\
&H_{\rm int}(Q)=-\frac{g}{2}Q^4+\frac{\lambda}{6}Q^6
\end{aligned}
\label{sym}
\end{equation}

The minima of the total potential in (\ref{sym}), $P=\omega_i^2Q_i^2-\frac{g}{2}Q^4+\frac{\lambda}{6}Q^6$, can be defined from the condition $\frac{\partial P}{\partial Q}=0$, giving for non-negative roots $Q_0=0$ and $Q_{\pm}=\sqrt{\frac{g}{\lambda}\pm\sqrt\frac{\omega_{\rm F}^2-\omega_i^2}{\lambda}}$, where

\begin{equation}
\omega_{\rm F}^2=\frac{g^2}{\lambda}
\label{omf}
\end{equation}

For frequency $\omega_i>\omega_{\rm F}$, there it only one real root $Q=0$ and one global minimum of $P$. For $\omega_i<\omega_{\rm F}$, there are three real roots, $Q_0$ and $Q_{\pm}$. $Q_0$ gives metastable local minimum, and $Q_{\pm}$ gives global minima on the ``mexican hat'' surface \cite{scirep3}. Tunneling of the pseudo-vacuum state $Q_0$ to the true vacuum state $Q_{\pm}$ results in {\it SO}(3) to {\it SO}(2) symmetry breaking, which by Goldstone theorem gives two massless modes $\psi_i^{2,3}$ and one massive mode $\psi_i^{1}$, where $\psi$ describes field excitations around the global minimum $\bar{Q}$: $Q=\bar{Q}+\psi$ \cite{bogol}. This means that for $\omega_i<\omega_{\rm F}$, one mode, $\psi_i^{1}$, contributes to the energy but two other modes, $\psi_i^{2,3}$, do not. For $\omega_i>\omega_{\rm F}$, there is no second minimum, associated symmetry breaking, and hence all three modes contribute to the energy \cite{scirep3}. Therefore, the Hamiltonian in terms of $\psi$ is

\begin{equation}
H=\frac{1}{2}\left(\sum_i\left(\dot{\psi_i^1}\right)^2+\sum_i\omega_i^2\left(\psi_i^1\right)^2\right)+
\frac{1}{2}\left(\sum_i\left(\dot{\psi_i}^{2,3}\right)^2+\sum_{\omega_i>\omega_{\rm F}}\omega_i^2\left(\psi_i^{2,3}\right)^2\right)
\label{psi}
\end{equation}
\noindent where the first two terms give the energy of the longitudinal mode and the last two terms give the energy of two transverse modes which has the potential energy of restoring forces at frequency $\omega_i>\omega_{\rm F}$ only.

Eq. (\ref{psi}) can be re-written in the form similar to Eq. (\ref{len}) where the first term corresponds to the total kinetic energy:

\begin{equation}
H=\frac{1}{2}\left(\sum_i\left(\dot{\psi_i}\right)^2+\sum_i\omega_i^2\left(\psi_i^1\right)^2+\sum_{\omega_i>\omega_{\rm F}}\omega_i^2\left(\psi_i^{2,3}\right)^2\right)
\label{psi1}
\end{equation}
\noindent

We have therefore shown that Hamiltonian (\ref{psi1}) is equivalent to the Hamiltonian of interacting field (\ref{sym}) or (\ref{field}) with a general multi-well interaction as in (\ref{int}). Importantly, Eq. (\ref{psi1}) includes potential energy of one mode $\psi^1$ and potential energies of two modes $\psi^{2,3}$ but with frequencies $\omega>\omega_{\rm F}$, i.e. gives the same energy as the energy of the liquid in Eqs. (\ref{len}-\ref{len1}). Hence, Eq. (\ref{psi1}) is equivalent to the Hamiltonian of the liquid. Therefore, the results for the liquid derived in the previous section apply to the interacting fields. This is the main idea of our proposed programme of treating interacting fields, which we develop in the next sections.

\section{Frenkel energy gap}

In the proposed ``liquid'' approach to interacting fields, the energy of the interacting field (\ref{field}) with the multi-well potential such as (\ref{int}), (\ref{sym}) or similar ones is the sum of the energy of motion in one well and the energy of inter-well motion, as is the case for the liquid Hamiltonian. At frequency larger than $\omega_{\rm F}$, no inter-well hopping motion takes place, and field $\phi$ oscillates in one harmonic well. At frequency smaller than $\omega_{\rm F}$, hopping takes place and results in the loss of restoring forces for $n-1$ modes with frequency $\omega<\omega_{\rm F}$ in Eq. (\ref{psi1}) (these are massless modes due to symmetry breaking discussed above whose loss is analogous to the loss of transverse modes in liquids with $\omega<\omega_{\rm F}$; $n$ is the number of field components).

This result has two implications. Lets consider the matter field $u$ and the gauge field $B$ that enforces the required symmetries of the problem \cite{bogol}:

\begin{equation}
\begin{aligned}
H(u,B)=H_0(u,D(B)u)+H_{\rm gauge}(B)\\
H_0=\frac{1}{2}\left(\left(D(B)u\right)^2+\mu^2u^2\right)+H_{\rm int}(u)
\label{matgauge}
\end{aligned}
\end{equation}
\noindent where $D(B)u$ is the covariant derivative related to the local symmetry of the Hamiltonian, $H_{\rm int}(u)$ is the interaction term, and where we continue operating in terms of Hamiltonians noting that the same results apply to the Lagrangian formulation. $u$ can be real or complex.

The {\it first} implication of our result is related to the matter field $u$. If $u$ is a strongly interacting field and $\frac{1}{2}\mu^2u^2+H_{\rm int}(u)$ in Eq. (\ref{matgauge}) has the form in Figure 2, our treatment enables us to straightforwardly calculate the energy of this interacting field according to Eq. (\ref{psi1}). This calculation involves no perturbation theory, associated divergences as well as non-renormalizability issues if $H_{\rm int}(u)$ contains powers higher than 4 as is the case in Eq. (\ref{sym}). This will be discussed in more detail in the next section. Importantly, the summation in the last term in Eq. (13) starts from frequency $\omega_{\rm F}$. Therefore, (\ref{psi1}) naturally predicts an energy gap in the spectrum below the ``Frenkel energy'', $E_{\rm F}$:

\begin{equation}
E_{\rm F}=\hbar\omega_{\rm F}
\label{frenergy}
\end{equation}

The appearance of $E_{\rm F}$ in the energy of the interacting field is a non-trivial result, yet it naturally emerges in our theory. This result is analogous to the appearance of the gap in the energy of two transverse modes in liquids discussed earlier.

We note that the energy gap emerges in both classical and quantum cases. In the latter case, quantization is performed at the level of the free field given by the first two terms in Eq. (\ref{field}), and the emergent $E_{\rm F}$ is related to the energy gap of the field quanta. This is analogous to the energy gap of transverse oscillation quanta in liquids, phonons, in Eq. (\ref{enf}).

It is interesting to compare our mechanism to the previous common approach aimed at finding a way to alter the mass of the field in order to enable the description of both massive and massless particles. In this approach, the second term in $H_0$ in Eq. (\ref{matgauge}) is negative ($\mu^2<0$), giving the imaginary mass (frequency) and implying the instability and absence of oscillating field quanta without $H_{\rm int}(u)$ \cite{bogol,peskin,ryder}, i.e. the absence of mass. Adding the interaction $H_{\rm int}(u)=hu^4$ stabilizes the field and gives two minima in the total potential $-\mu^2u^2+hu^4$. Symmetry breaking $u=u_0+v$ around the new minimum $u_0$ results in one component of $v$ acquiring the mass and $n-1$ components remaining massless \cite{bogol}. If the interaction of the initially massless scalar field $u$ (if $\mu^2<0$ is assumed) and the initially massless gauge field such as the Yang-Mills field is added, the symmetry breaking results in the gauge field acquiring the mass via the Higgs effect and initially massless scalar field $u$ acquiring mass and giving rise to the massive Higgs boson. Similarly negative $\mu^2$ features in other considerations including the Weinberg-Salam model: the model assumes two initially massless gauge fields which acquire mass via the Higgs effect by introducing an auxiliary scalar field with negative $\mu^2$, with concomitant acquiring of mass by the scalar field \cite{bogol}.

An important reason for writing the potential as $-\mu^2u^2+hu^4$ with the negative first term is that it is the simplest positively-defined form that gives two vacuum states with the associated symmetry breaking and, importantly, contains the quadric interaction only, which is renormalizable. If the positive first term $\mu^2u^2$ is used instead, as is the case in general schemes of field quantization \cite{bogol}, the existence of more than one (symmetric) minimum to enable symmetry breaking requires powers of $u$ higher than 4 as in, for example, Eq. (\ref{sym}). This makes perturbation theory non-renormalizable.

Notably, renormalizability is irrelevant in our approach because we do not use perturbation theory with associated divergences. Therefore, our approach contains positive $\omega_i^2, \mu^2>0$ and at the same time has the capacity to alter the mass of the field so that both massive and massless particles appear as a result of interaction in the same field. This point is discussed below in more detail.

Consequently, the important feature of our mechanism is the positivity of the second term in Eqs. (\ref{field}), (\ref{sym}), (\ref{psi1}), and (\ref{matgauge}): $\omega_i^2, \mu^2>0$. This gives real masses (frequencies) in the harmonic non-interacting case and enables $u$ to oscillate in stable harmonic minima, i.e. provides a physically sensible starting point for the field as the general quantization scheme of fields demands \cite{bogol}. If an interaction is added so that the total potential is multi-well as shown in Figure 2, the field $u$ continues to oscillate in a single minimum, but also acquires the ability to hop to other minima. This hopping modifies the spectrum in such a way that oscillations of $n-1$ modes are maintained at frequency $\omega>\omega_{\rm F}$, but are lost at frequency $\omega<\omega_{\rm F}$, giving the Frenkel energy gap, $E_{\rm F}$. As discussed in the previous section, this modification involves symmetry breaking as in the Higgs effect although the physical essense of our mechanism is notably different.

The {\it second} implication of our result is related to the gauge field $H_{\rm gauge}(B)$. If $H_{\rm gauge}(B)$ is a strongly interacting field, has powers of the field higher than 2, and can generally be represented by the potential of the form in Figure 2, our approach readily predicts an energy gap in the spectrum given by $E_{\rm F}$ in Eq. (\ref{frenergy}), as it does for the matter field discussed above. Notably, the emergence of the energy gap in our approach is the property of the anharmonic interacting field itself, be it gauge field or matter field.

We consequently propose that the energy gap in Eq. (\ref{frenergy}) is relevant for the currently important problem of demonstrating the existence of the mass gap in the anharmonic gauge theory containing fields to order higher than two (such as, for example, the Yang-Mills theory \cite{new6,new7,new8}). As discussed above, this theory can be classical or quantum. Our demonstration does not require proving the renormalizability of the interacting anharmonic theory because we do not employ the perturbation approach with associated divergences.

Gapped field excitations are related to short-range forces with massive particles. Then, the mass of the short-range ``Frenkel particle'' associated with $E_{\rm F}=\hbar\omega_{\rm F}$ is

\begin{equation}
m_{\rm F}=\frac{\hbar\omega_{\rm F}}{c^2}
\label{frmass}
\end{equation}

The Frenkel particle is therefore a short-ranged massive particle naturally emerging in our approach. The Frenkel particle emerges from introducing interaction in the field and resulting disappearance of $n-1$ ``transverse modes'' at frequency $\omega<\omega_{\rm F}$ due to the Frenkel mechanism discussed above.

For $n$-component field, there are $n-1$ massive Frenkel particles, with $E_{\rm F}$ and $m_{\rm F}$ of each particle defined by the shape of the multi-well potential in $n$ dimensions. In general, the shape of this potential can vary for each field component, yielding different $\omega_{\rm F}$ (see Eq. (\ref{omf})) and different $E_{\rm F}$ and $m_{\rm F}$ as a result.

We further observe that in addition to the Frenkel energy gap for $n-1$ ``transverse'' modes, the energy spectrum of the anharmonic field Hamiltonian (\ref{psi1}) also contains one gapless ``longitudinal'' mode, analogous to one gapless longitudinal mode in liquids. Gapless field excitations are related to long-range forces with massless particles. Therefore, both short-range forces with massive particles and long-range forces with massless particles naturally emerge and {\it unify} in a single interacting field, an interesting result not hitherto anticipated.

$\omega_{\rm F}$ and $m_{\rm F}$ are determined by the interaction potential. In liquids, $\omega_{\rm F}$ is limited by the largest Debye frequency, $\omega_{\rm D}$, corresponding to the shortest wavelength comparable to the interatomic separation. In field theory, the upper limit for $\omega_{\rm D}$ (and hence $\omega_{\rm F}$) is tentatively given by the shortest wavelength comparable to the Planck length. Then, the upper limit for $m_{\rm F}$ is given by the Planck mass.

\section{Interacting fields: energy}

The Hamiltonian of the free field such as the scalar Klein-Gordon field can be written in terms of harmonic excitation quanta as $H=E_0+\sum\limits_k n_k\hbar\omega_k$, where $n_k$ is the number of excitations and $E_0=\frac{1}{2}\sum\limits_k\hbar\omega_k$ is the ground-state (vacuum) energy. As discussed above, the energy of the interacting field is given by Eq. (\ref{psi1}). Its liquid equivalent is Eq. (\ref{len}) which can be re-written as Eq. (\ref{len1}). Then, the Hamiltonian of the interacting 3-component field can be re-written in the same way:

\begin{equation}
H=E_0+\sum\limits_{\omega_k}n_k\hbar\omega_k^l+\sum\limits_{\omega_k>\omega_{\rm F}}n_k\hbar\omega_k^t+\frac{1}{2}\sum\limits_{\omega_k<\omega_{\rm F}}n_k\hbar\omega_k^t
\label{fienergy}
\end{equation}
\noindent where $\omega^l$ and $\omega^t$ correspond to longitudinal and transverse modes, respectively.

$\sum\limits_{\omega_k}n_k\hbar\omega_k^l+\sum\limits_{\omega_k>\omega_{\rm F}}n_k\hbar\omega_k^t$ give the energy of one longitudinal mode and two transverse modes with frequency $\omega_k>\omega_{\rm F}$. As in Eq. (\ref{len1}), the last term in Eq. (\ref{fienergy}) ensures that the loss of two transverse modes with frequency $\omega_k<\omega_{\rm F}$ is accompanied by the reduction of their potential energy only (as Eq. (\ref{psi1}) requires), but not by the reduction of the total kinetic energy in Eq. (\ref{psi1}).

For the free field, the zero-point oscillatory energy $E_0$ is ignored for the reason that experimental quantities are related to the difference between the measured energy and $E_0$ \cite{peskin}. For liquids, we ignored $E_0$ when calculating $c_v$ from Eq. (\ref{enf}) because this energy was small compared to the range of liquid temperatures (see Figure 1).

For the interacting field, $E_0$ is the sum of zero-point energies of one longitudinal modes and two transverse modes with $\omega_k>\omega_{\rm F}$:

\begin{equation}
E_0=\frac{1}{2}\left(\sum\limits_{\omega_k}\hbar\omega_k^l+\sum\limits_{\omega_k>\omega_{\rm F}}\hbar\omega_k^t\right)
\label{vacuum}
\end{equation}

$E_0$ can be evaluated using the density of states which, for illustration purposes, we take in the quadratic Debye form, $g(\omega)\propto\omega^2$, associated with $\omega\propto k$ dependence. Sums over $\omega_k$ in Eq. (\ref{vacuum}) can be evaluated as $\int\limits_0^{\omega_{\rm D}}\hbar\omega g_l(\omega)d\omega$ and $\int\limits_{\omega_{\rm F}}^{\omega_{\rm D}}\hbar\omega g_t(\omega)d\omega$, respectively, where $g_l=\frac{3\omega^2}{\omega_{\rm D}^3}$ and $g_l=\frac{6\omega^2}{\omega_{\rm D}^3}$ are densities of states of longitudinal and transverse modes per oscillator \cite{prb1}. This gives the zero-point energy of the liquid and of the 3-component interacting field:

\begin{equation}
E_0=\frac{3}{8}\hbar\omega_{\rm D}\left(3-2\left(\frac{\omega_{\rm F}}{\omega_{\rm D}}\right)^4\right)
\label{zero1}
\end{equation}

We therefore find that interaction leads to lowering of the vacuum energy by amount $\frac{3}{4}\hbar\omega_{\rm D}\left(\frac{\omega_{\rm F}}{\omega_{\rm D}}\right)^4$ ($\omega_{\rm F}=0$ corresponds to the non-interacting case).

We note that for illustration purposes, we have used the quadratic Debye density of states to calculate the energy. Depending on the equations of motion of the particular field considered, $g(\omega)$ can be different. For example, in the case of the Klein-Gordon field where $\omega^2=k^2+m^2$, $g(\omega)$ shifts, and can be approximately written as $g(\omega)\propto(\omega-m)^2$. In this case, additional terms appear in Eqs. (\ref{zero1}) with lower powers of $\frac{\omega_{\rm F}}{\omega_{\rm D}}$.

Eq. (\ref{zero1}) can be generalized for the $n$-component interacting fields. In this case, $g_l(\omega)$ and $g_t(\omega)$ modify as $g_l(\omega)=\frac{n\omega^{n-1}}{\omega_{\rm D}^n}$ and $g_t(\omega)=\frac{n(n-1)\omega^{n-1}}{\omega_{\rm D}^n}$, giving

\begin{equation}
E_0=\frac{n}{2(n+1)}\hbar\omega_{\rm D}\left(n-(n-1)\left(\frac{\omega_{\rm F}}{\omega_{\rm D}}\right)^{n+1}\right)
\label{zero2}
\end{equation}

Eqs. (\ref{zero1}-\ref{zero2}) serve to illustrate several points. $\tau\gg\tau_{\rm D}$ ($\omega_{\rm F}\ll\omega_{\rm D}$) corresponds to the ``rare hopping'' regime when field $\phi$ in Eq. (\ref{field}) (or (\ref{sym})) oscillates in one potential minimum many times before hopping to a nearby minimum, similar to the dynamics of viscous liquids discussed in Section 2. In this regime, the second term in brackets in Eqs. (\ref{zero1}) and (\ref{zero2}) is close to zero, giving the same zero-point energy as in the non-interacting case. $E_0$ appreciably changes in the ``frequent hopping'' regime only, when $\frac{\omega_{\rm F}}{\omega_{\rm D}}\approx 1$ and the second term in brackets Eqs. (\ref{zero1}-\ref{zero2}) is close to 1, similar to what takes place for the liquid energy in Eqs. (\ref{class}-\ref{enf}).

When $\frac{\omega_{\rm F}}{\omega_{\rm D}}\approx 1$, $E_0=\frac{3}{8}\hbar\omega_{\rm D}$ in Eq. (\ref{zero1}) and $E_0=\frac{n}{2(n+1)}\hbar\omega_{\rm D}$ in Eq. (\ref{zero2}). This reduces the zero-point energy by a factor of 3 and $n$, respectively. We therefore find that depending on $\omega_{\rm F}$ set by Eqs. (\ref{sym}-\ref{omf}), the vacuum energy can vary significantly, an interesting result relevant for cosmology research where the relationship between the cosmological constant and the vacuum energy is discussed \cite{peskin}.

We note that significant changes of the energy due to interaction take place only when $\frac{\omega_{\rm F}}{\omega_{\rm D}}\approx 1$ means that our approach is essentially {\it non-perturbative}. Indeed, the interaction strength can be quantified by $\frac{\omega_{\rm F}}{\omega_{\rm D}}$ ($\omega_{\rm F}=0$ corresponds to the absence of interaction and hopping). Therefore, significant changes of the energy due to interaction take place only when the interaction strength is not small. Consequently, Eqs. (\ref{zero1}-\ref{zero2}) can not be obtained in perturbation theory to any finite order.

We also note that the upper integration limit used to calculate (\ref{zero1}) is given by finite $\omega_{\rm D}$, avoiding the ultraviolet divergence. For a quantized field, $\omega_{\rm D}$ corresponds to the largest frequency of oscillation in a single-well potential in Fig. 2, or to the shortest wavelength discussed in the previous section. A discussion of using upper integration limits in statistical physics and field theory problems is reviewed in Ref. \cite{berges-review}.

%It is interesting to ask how the energy changes when $\omega_{\rm F}$ starts to exceed $\omega_{\rm D}$. In liquids, $\omega_{\rm F}=\omega_{\rm D}$ gives the energy crossover on temperature increase, corresponding to the Frenkel crossover \cite{phystoday,pre,prl} discussed above. Further increase of $\omega_{\rm F}$ results in the disappearance of the remaining longitudinal mode \cite{natcom}. This disappearance starts with the short wavelength equal to the particle mean free path that increases with temperature (note that this is contrary to the decrease of transverse modes in the regime $\omega_{\rm F}<\omega_{\rm D}$, which disappear starting with the longest wavelength). In the high-temperature limit, all oscillatory modes in the liquid are lost, corresponding to the ideal gas with specific heat of $\frac{3}{2}$ \cite{natcom}.

%Similarly, when the longitudinal mode of the field disappears in the ``high-temperature'' limit, the motion of $\phi$ is purely kinetic and ballistic with no oscillations at any frequency, corresponding to the last two terms in Eq. (\ref{psi}) becoming zero. This implies that the vacuum energy in Eq. (\ref{vacuum}) becomes zero. Therefore, depending on the form of the potential and dynamics of the field, the vacuum energy can vary between zero and its maximal value, an interesting result relevant for cosmology research where the relationship between the cosmological constant and the vacuum energy is discussed \cite{peskin}.

\section{Interacting fields: propagator}

The quantities of interest in the quantum field theory and particle physics include the propagator, $K(x,x^\prime,t)=\langle x^\prime|\exp(-iHt)|x\rangle=\int Dx\exp{\frac{iS}{\hbar}}$. When $H=H_0+H_{\rm int}$, where $H_{\rm int}$ is the interaction, the existing method to calculate $K(x,x^\prime,t)$ is by the perturbative expansion of the path integral in terms of $H_{\rm int}$ and handling Feynman diagrams in the expansion.

In our non-perturbative approach to interacting field theory, we calculate the propagator for the strongly interacting field on the basis of two observations. First, $K(x,x^\prime,t)$ is identical to the density matrix $\rho(x,x^\prime,T)=\frac{\exp\left(-\frac{H}{T}\right)}{Z}$ in statistical physics, where $Z$ is the partition function, on substitution $T=\frac{\hbar}{it}$ \cite{feynman}. $\rho(x,x^\prime,T)$ for the normal mode of the quantum harmonic oscillator with frequency $\omega$ is (see, e.g., \cite{feynman}):

\begin{equation}
\rho_\omega(x,x^\prime,T)=\left(\frac{\omega}{2\pi\hbar\sinh 2f}\right)^{\frac{1}{2}}\exp\left(\frac{-\omega}{2\hbar\sinh 2f}\left((x^2+{x^\prime}^2)\cosh 2f-2xx^\prime\right)\right)
\label{prop1}
\end{equation}
\noindent where $f=\frac{\hbar\omega}{2T}$.

For $x=x^\prime$, $\rho$ simplifies to

\begin{equation}
\rho_\omega(x,x,T)=\left(\frac{\omega}{2\pi\hbar\sinh 2f}\right)^{\frac{1}{2}}\exp\left(-\frac{\omega}{\hbar}x^2\tanh f\right)
\label{prop2}
\end{equation}

Eq. (\ref{prop2}) is the unnormalized Gibbs probability distribution for the harmonic oscillator \cite{lanstat}.

Second, if the Hamiltonian of the system is the sum of energies of normal modes, the probability distribution for the whole system is the product of distributions for individual normal modes \cite{lanstat}. As we have shown above, the Hamiltonian of the strongly-interacting field moving in the multi-well potential is given by Eq. (\ref{psi1}), and the energy by Eq. (\ref{len1}). Therefore, the density matrix of the field becomes the product of density matrices given by Eqs. (\ref{prop1}) or (\ref{prop2}), and the terms in the product are the modes that contribute to the total energy of the interacting field in Eq. (\ref{fienergy}) or its liquid equivalent, Eq. (\ref{len1}).

For simplicity, we consider $x=x^\prime$ and $\rho_\omega(x,x,T)$ in Eq. (\ref{prop2}) above, with the corresponding Feynman propagator $K_\omega(x,t)$ obtained by putting $T=\frac{\hbar}{it}$. Then,

\begin{equation}
K_\omega(x,t)=\left(\frac{\omega}{2\pi\hbar i\sin\omega t}\right)^{\frac{1}{2}}\exp\left(-\frac{i}{\hbar}\omega x^2\tan\frac{\omega t}{2}\right)
\end{equation}

and the propagator of the strongly-interacting field, $K(x,t)$, is

\begin{equation}
K(x,t)=\prod\limits_\omega K_\omega(x,t)\\
\end{equation}
\noindent where the product is over the modes that give the total energy of the interacting field in Eq. (\ref{fienergy}) or Eq. (\ref{len1}).

We consider two limiting cases: classical high-temperature limit $\frac{\hbar\omega}{T}\ll 1$ (corresponding to $\omega t\ll 1$ and short-time regime) and strongly quantum limit $\frac{\hbar\omega}{T}\gg 1$ (corresponding to $\omega t\gg 1$ and long-time regime). In the first case, $\rho_\omega(x,x,T)$ in Eq. (\ref{prop2}) is $\rho_\omega(x,x,T)=\left(\frac{T}{2\pi\hbar^2}\right)^{\frac{1}{2}}\exp\left(-\frac{\omega^2x^2}{2T}\right)$ (here, $\exp\left(-\frac{\omega^2x^2}{2T}\right)$ is the unnormalized probability distribution of the classical harmonic oscillator), $K_\omega(x,t)=\left(\frac{1}{2\pi\hbar i t}\right)^{\frac{1}{2}}\exp\left(-\frac{i\omega^2x^2t}{2\hbar}\right)$, and

\begin{equation}
K(x,t)=\left(\prod\limits_i\left(\frac{1}{2\pi\hbar i t}\right)^{\frac{1}{2}}\right)\exp\left(-\frac{ix^2t}{2\hbar}\sum\limits_i\omega_i^2\right)
\label{prophigh}
\end{equation}

The sum in Eq. (\ref{prophigh}) is over the classical oscillator energies (each energy is $\propto\omega^2$). The sum includes all modes that contribute to the total energy of the interacting field. According to Eq. (\ref{len1}) or (\ref{fienergy}), this energy is $E_l+E_t(\omega>\omega_{\rm F})+\frac{1}{2}E_t(\omega<\omega_{\rm F})$. As before, these terms can be evaluated as $\int\limits_0^{\omega_{\rm D}}\omega^2 g_l(\omega)d\omega$, $\int\limits_{\omega_{\rm F}}^{\omega_{\rm D}}\omega^2g_t(\omega)d\omega$ and $\frac{1}{2}\int\limits_0^{\omega_{\rm F}}\omega^2 g_t(\omega)d\omega$ with Debye density of states $g_l=\frac{3P\omega^2}{\omega_{\rm D}^3}$ and $g_t=\frac{6P\omega^2}{\omega_{\rm D}^3}$ (for the three-component field and with the same proviso of using the Debye model as discussed in the previous section), where $P$ is the number of oscillators in one mode. The sum becomes $\frac{3}{5}P\omega_{\rm D}^2\left(3-\left(\frac{\omega_{\rm F}}{\omega_{\rm D}}\right)^5\right)$. The number of terms in the product in Eq. (\ref{prophigh}) can be similarly estimated as the sum of $\int\limits_0^{\omega_{\rm D}}g_l(\omega)d\omega$, $\int\limits_{\omega_{\rm F}}^{\omega_{\rm D}}g_t(\omega)d\omega$ and $\frac{1}{2}\int\limits_0^{\omega_{\rm F}}g_t(\omega)d\omega$, giving  $P\left(3-\left(\frac{\omega_{\rm F}}{\omega_{\rm D}}\right)^3\right)$ (compare with Eq. \ref{class}). Then,

\begin{equation}
K(x,t)=\left(\frac{1}{2\pi\hbar i t}\right)^{\frac{P}{2}\left(3-\left(\frac{\omega_{\rm F}}{\omega_{\rm D}}\right)^3\right)}
\exp\left(-\frac{3P\omega_{\rm D}^2}{10}\left(3-\left(\frac{\omega_{\rm F}}{\omega_{\rm D}}\right)^5\right)\frac{ix^2t}{\hbar}\right)
\label{prophigh1}
\end{equation}

In the strongly quantum limit $\frac{\hbar\omega}{T}\gg 1$, or long-time regime $\omega t\gg 1$, Eq. (\ref{prop2}) becomes $\rho_\omega(x,x,T)=\left(\frac{\omega}{\pi\hbar\exp\frac{\hbar\omega}{T}}\right)^{\frac{1}{2}}\exp\left(-\frac{\omega x^2}{\hbar}\right)$ (here, $\exp\left(-\frac{\omega x^2}{\hbar}\right)$ is the square of the amplitude of the wave function in the ground state of the normal mode), $K_\omega(x,t)=\frac{\omega}{\pi\hbar}\exp(-i\omega t)\exp\left(-\frac{\omega x^2}{\hbar}\right)$, and

\begin{equation}
K(x,t)=\left(\prod\limits_i\frac{\omega_i}{\pi\hbar}\right)\exp\left(-\left(it+\frac{x^2}{\hbar}\right)\sum\limits_i\omega_i\right)
\label{proplow}
\end{equation}

As in Eq. (\ref{prophigh}), the sum in Eq. (\ref{proplow}) is over modes contributing to the total energy of the interacting field, $E_l+E_t(\omega>\omega_{\rm F})+\frac{1}{2}E_t(\omega<\omega_{\rm F})$, and can be evaluated in the same way as before, giving $\frac{3}{4}P\omega_{\rm D}\left(3-\left(\frac{\omega_{\rm F}}{\omega_{\rm D}}\right)^4\right)$. The number of terms in the product in Eq. (\ref{proplow}) is the same as in Eq. (\ref{prophigh}), and is $P\left(3-\left(\frac{\omega_{\rm F}}{\omega_{\rm D}}\right)^3\right)$. Finally, introducing the average geometric frequency, $\bar\omega$, as $\prod\limits_i\omega_i=\bar\omega^{P\left(3-\left(\frac{\omega_{\rm F}}{\omega_{\rm D}}\right)^3\right)}$, Eq. (\ref{proplow}) becomes

\begin{equation}
K(x,t)=\left(\frac{\bar{\omega}}{\pi\hbar}\right)^{P\left(3-\left(\frac{\omega_{\rm F}}{\omega_{\rm D}}\right)^3\right)}
\exp\left(-\frac{3P\omega_{\rm D}}{4}\left(3-\left(\frac{\omega_{\rm F}}{\omega_{\rm D}}\right)^4\right)\left(it+\frac{x^2}{\hbar}\right)\right)
\label{proplow1}
\end{equation}

Eqs. (\ref{prophigh}-\ref{proplow1}) can be generalized to the case of $n$-component field in the same way as in the calculation of the energy in the previous section.

Similarly to the energy discussed in the previous sections, $K(x,t)$ in Eqs. (\ref{prophigh1}) and (\ref{proplow1}) remain close to their non-interacting values when $\frac{\omega_{\rm F}}{\omega_{\rm D}}\ll 1$, corresponding to the rare hopping regime, the regime where the field oscillates many time in one minimum of the multi-well potential before hopping to the next. Only when $\frac{\omega_{\rm F}}{\omega_{\rm D}}\approx 1$ in the frequent hopping regime, factors such as $3-\left(\frac{\omega_{\rm F}}{\omega_{\rm D}}\right)^l$ reduce noticeably and significantly change $K(x,t)$, similarly to what takes place for the energy in Eqs. (\ref{zero1}-\ref{zero2}) as well as Eqs. (\ref{class}-\ref{enf}).

Similarly to the energy, the last observation that the important changes of $K(x,t)$ due to interaction take place only when $\frac{\omega_{\rm F}}{\omega_{\rm D}}\approx 1$ implies that these changes can not be derived using the perturbation approach by expanding in terms of the interaction strength $\frac{\omega_{\rm F}}{\omega_{\rm D}}$. Furthermore, even if $\frac{\omega_{\rm F}}{\omega_{\rm D}}$ is small, the propagators in Eqs. (\ref{prophigh1}) and (\ref{proplow1}) can not be expanded for large $x$ and $t$ because the argument in the exponent is large. Hence, Eqs. (\ref{prophigh1}) and (\ref{proplow1}) can not be obtained in perturbation theory to any finite order, emphasizing the non-perturbative nature of our approach.

%The effects of interaction on the relativistic propagator, $K(x)=\frac{1}{(2\pi)^4}\int d^4k\frac{\exp(-ikx)}{k^2-m^2}$, can be incorporated in the way similar to the calculation above: the integral becomes the sum of three integrals with different integration limits: the first integrates over one unmodified longitudinal mode, the second integrates over transverse modes from $\omega_{\rm F}$ to $\omega_{\rm D}$ and the third one integrates over transverse modes from zero to $\omega_{\rm F}$ (with a factor of $\frac{1}{2}$ in front of the sum over energies).

The above calculation serves two purposes. First, it illustrates the scheme of evaluating propagators in the proposed non-perturbative approach to interacting fields. Importantly, $K(x,t)$ can be non-perturbatively calculated in closed form for any shape of the multi-well potential in Figure 2, including for $H_{\rm int}(\phi)=-\frac{g}{2}\phi^4+\frac{\lambda}{6}\phi^6$ in Eqs. (\ref{int}-\ref{sym}) for which the perturbation theory is non-renormalizable as well as for $H_{\rm int}(\phi)$ containing higher powers of $\phi$. Second, it demonstrates how $\omega_{\rm F}$ emerges in the propagator (see Eqs. (\ref{prophigh1}), (\ref{proplow1})). This will remain to be the case if more complicated fields are considered or interactions with other fields are taken into account, implying that the Frenkel energy will feature in the results and the Frenkel particle will emerge and interact with other particles and fields.

\section{Relationship to the hydrodynamic description of quark-gluon plasma}

It is interesting to compare our approach to that involved in deriving temperature dependence of viscosity of the quark-gluon plasma \cite{poli1,poli2}. Describing the plasma by the strongly-interacting $N=4$ supersymmetric Yang-Mills theory and invoking the AdS/CFT correspondence, the authors derive the cubic temperature-dependence of viscosity of the plasma, $\eta\propto T^3$, and further assert that viscosity is proportional to entropy $S$, $\eta\propto S$ \cite{poli1,poli2}, on the basis of the earlier result $S\propto T^3$ \cite{gubser}. Our Hamiltonians describing liquids in condensed matter (\ref{field}--\ref{psi1}) are different, precluding the direct comparison of results. However, we make two observations regarding the two approaches.

First, there are two regimes in which liquid dynamics operates: the commonly considered hydrodynamic regime where $\omega\tau<1$ and the solid-like elastic regime $\omega\tau>1$ \cite{frenkel}. The two regimes are described by different equations and have different properties including dissipation laws \cite{frenkel}. The approach in Refs. \cite{poli1,poli2} is based on the hydrodynamic description, and discusses hydrodynamic modes that generally emerge in the long-range and low-frequency behavior of any interacting system. Notably, our approach is based on modes with frequency $\omega>\frac{1}{\tau}$, modes that propagate in the solid-like elastic regime. The importance of these modes is evident from their dominant contribution to the energy due to $g(\omega)\propto\omega^2$. Indeed, non-hydrodynamic solid-like elastic modes, both longitudinal and transverse, with frequency close to $\omega_{\rm D}$ give most of the liquid energy, as they do in a solid. This is emphasized by, for example, Eq. (\ref{class}) which gives the solid-like energy $\epsilon=3T$ unless $\omega_{\rm F}$ starts to approach $\omega_{\rm D}$. It is therefore important to include these non-hydrodynamic modes in the consideration of any strongly-interacting theory.

Second, we find that despite the difference in starting Hamiltonians, our approach predicts similar relationship $S\propto T^3$, albeit on the basis of a different mechanism. Indeed, Eq. (\ref{enf}) gives in the quantum low-temperature limit ($D(x)=\frac{\pi^4}{5x^3}$ for $x\gg 1$):

\begin{equation}
\epsilon=\epsilon_0+\frac{2\pi^4}{5}\frac{T^4}{\left(\hbar\omega_{\rm D}\right)^3}
\end{equation}
\noindent

Interestingly, $\epsilon$ does not depend on $\omega_{\rm F}$ which cancels out in Eq. (\ref{enf}). This implies that transverse modes do not contribute to the liquid energy in the quantum low-temperature limit: as discussed earlier \cite{prb1}, only low-frequency phonons can be excited and contribute to the energy in this limit, but low-frequency transverse phonons are not propagating in the liquid \cite{frenkel}. Then, $c_v=\frac{8\pi^4}{5}\frac{T^3}{\left(\hbar\omega_{\rm D}\right)^3}$ and specific entropy, $s=\int\frac{c_v}{T}dT$, is:

\begin{equation}
s=\frac{8\pi^4}{15}\frac{T^3}{\left(\hbar\omega_{\rm D}\right)^3}
\end{equation}
\noindent Similar $s\propto T^3$ dependence is obtained if one considers the energy of the liquid in the quantum gas-like regime above the Frenkel line \cite{natcom}.

We therefore find that similar to Ref. \cite{gubser}, the entropy increases as $T^3$, the result that can be attributed to the similarity of statistical distributions of the gas of excitations considered: brane excitations in Ref. \cite{gubser} and phonon excitations in this work.

Interestingly different from the previous result \cite{poli1,poli2,gubser}, the entropy is not proportional to viscosity in our approach because viscosity of liquids does not follow the relationship $\eta\propto T^3$ predicted for the quark-gluon plasma. Indeed, experimental and theoretical viscosity of liquids described by our condensed-matter Hamiltonians is known to have two regimes: liquid-like regime where $\eta$ decreases with temperature either exponentially or faster via the Vogel-Fulcher-Tammann law ($\eta\propto\exp\left(\frac{A}{T-T_0}\right)$, where $A$ and $T_0$ are constants) and gas-like regime \cite{phystoday,pre,prl} where $\eta$ increases with temperature as $\eta\propto T^\gamma$ with $\gamma\approx 0.5-0.6$ \cite{natcom} (for the ideal non-interacting gas, $\gamma=0.5$).

We therefore find interesting similarities and differences between the two approaches that warrant further discussion.

\section{Summary and outlook}

We have outlined a programme of treating interacting fields where the interaction strength is arbitrarily high and is generally represented by a multi-well interaction potential with powers higher than two. The programme does not involve perturbation theories and associated problems related to divergences. In this approach, the motion of the field consists of oscillations in one harmonic well and hopping between different wells. Accounting for this motion results in the disappearance of the ($n$-1) modes with frequency smaller than the Frenkel frequency $\omega_{\rm F}$, similarly to the loss of two transverse modes in a liquid with frequency $\omega<\omega_{\rm F}$. We have illustrated the proposed programme with the calculation of the energy and propagator of the strongly-interacting field. We have shown that the energy and the propagator of the interacting field are equal to their non-interacting values (i.e. when the potential term includes the quadratic term only) in the rare-hopping regime, but noticeably change in the frequent hopping regime, and emphasized that the results can not be obtained in perturbation theory.

In the proposed programme, the Frenkel energy gap for transverse modes, $E_{\rm F}=\hbar\omega_{\rm F}$, and associated massive ``Frenkel'' particle naturally appear in our consideration, the result that is relevant for current efforts to demonstrate a mass gap in interacting field theories such as Yang-Mills theory. Notably, emergence of the massive particle in this picture involves a physically sensible starting point in terms of real masses (frequencies) in the harmonic non-interacting field, in contrast to the Higgs effect involving the imaginary mass as a starting point. We further noted that the longitudinal mode remains gapless, implying that our treatment of the interacting field theory naturally gives rise to and unifies short-range and long-range forces with massive and massless particles, an important result not hitherto anticipated.

The similarity of problems and equations describing liquids and objects in the micro- and mega-world is not coincidental. In all cases, the theory often has no small parameter. In all cases, most interesting effects are related to the strong contribution of non-linear terms which can not be adequately described by the perturbation theory. In field theories, this leads to divergent series, infinities and associated problems. Our ability to describe many liquid properties by introducing the frequency of inter-well hopping suggests that this approach can advance the problems in other fields of physics discussed here, including those in micro- and mega-world. Indeed, physical phenomena in these worlds involve Hamiltonians that are formally similar to those discussed here including, for example, sigma models where effective approaches have been explored \cite{n1,n2,n3,n4,n5,new9,new6,new7,new8}.

We have noted marked changes in the energy and the propagator when $\omega_{\rm F}=\omega_{\rm D}$. This equality marks the Frenkel crossover in liquids \cite{phystoday,pre,prl,natcom} at which all major system properties undergo qualitative changes. A similar crossover should exist in the field theory with the interaction of the form shown in Figure 2, the point that warrants further discussion.

K. T. is grateful to L. H. Ryder, J. Sonner, P. Millington, A. Brandhuber, G. Travaglini, B. Spence, D. Bolmatov and E. Musaev for discussions and to EPSRC and SEPnet for support.

\end{document}